\documentstyle [12pt,a4p,epsfig,amsmath,multicol]{article}
\begin{document}
\vspace*{0.6cm}

\begin{center} 
{\normalsize\bf Quantum Interference Effects in the Detection
 Probability of Charged Leptons Produced in Charged Current Weak Interactions}
\end{center}
\vspace*{0.6cm}
\centerline{\footnotesize J.H.Field}
\baselineskip=13pt
\centerline{\footnotesize\it D\'{e}partement de Physique Nucl\'{e}aire et 
 Corpusculaire, Universit\'{e} de Gen\`{e}ve}
\baselineskip=12pt
\centerline{\footnotesize\it 24, quai Ernest-Ansermet CH-1211Gen\`{e}ve 4. }
\centerline{\footnotesize E-mail: john.field@cern.ch}
\baselineskip=13pt
 
\vspace*{0.9cm}
\abstract{The phases of quantum interference effects in charged lepton
  production by neutrinos (neutrino oscillations) following pion decay at
  rest and in flight, muon decay and 
  nuclear $\beta-$decay at rest as well as for `muon oscillations'
  following pion decay at rest and in flight are calculated. 
 The same phase is found for neutrino and muon oscillations following 
   pion decay at rest.
 The results found
  disagree with the conventionally used value: $\phi_{12} = \Delta m^2 L/(2P)$.
  Differences to previous treatments are briefly commented on.}
\vspace*{0.9cm}
\normalsize\baselineskip=15pt
\setcounter{footnote}{0}
\renewcommand{\thefootnote}{\alph{footnote}}
\newline
PACS 03.65.Bz, 14.60.Pq, 14.60.Lm, 13.20.Cz 
\newline
{\it Keywords ;} Quantum Mechanics,
Neutrino Oscillations.
\newline

\vspace*{0.4cm}

 This letter sketches briefly a derivation of spatially dependent interference
 effects (conventionally called `neutrino oscillations')
 in the detection probability of charged leptons produced in
  neutrino interactions following the
  weak decays of unstable `source' particles: pions, muons and $\beta-$radioactive
 nuclei. Similar effects for the detection probability of the decay 
  products of muons produced in pion decay (`muon oscillations') are 
  also considered.  The calculations are based on Feynman's path amplitude
  formulation of quantum
 mechanics~\cite{Feyn2} in which the probability of transition from a
 set of initial states $I = \sum_l i_l$ to a set of final states $F = \sum_m f_m$
 is given by the relation:
  \begin{equation}
 P_{FI} = \sum_m \sum_l \left|\sum_{k_1} \sum_{k_2}...\sum_{k_n}\langle f_m| k_1 \rangle
\langle k_1| k_2 \rangle...\langle k_n| i_l \rangle \right|^2  
 \end{equation}
 where  $k_j,~j=1,n$ are (unobserved) intermediate quantum states.
 Only the essential elements of the calculation and the results are presented
 here. Full details are given elsewhere~\cite{JHF}.
 \par The application of 
  Eqn(1) to pion decay at rest is illustrated in the ideal experiment shown in
  Fig.1.  A $\pi^+$ comes to rest in a stopping target T at time $t_0$ as recorded
  by the counter C$_{\rm A}$ (Fig.1a)). The pion at rest constitutes the initial state
  of the path amplitudes.
  In Fig.1b) and Fig.1c) are shown two alternative histories of the stopped pion.
  In Fig.1b) the pion decays at time $t_1$ into the neutrino mass eigenstate
  $|\nu_1>$, of mass $m_1$, and in Fig.1c) into the neutrino mass eigenstate
  $|\nu_2>$, of mass $m_2$, at the later time $t_2$.  If $m_1>m_2$, then, for a 
  suitable choice of the times $t_1$ and $t_2$, interference between the path amplitudes
  corresponding to the different physical processes shown in  Fig.1b) and Fig.1c)
  will occur when a neutrino interaction $(\nu_1,\nu_2) n \rightarrow e^- p$ takes place at time
  $t_D$ and distance $L$ from the pion decay point(Fig.1d)). The final state of the neutrino
  interaction event is also that of the path amplitudes.
  For this experiment, the path amplitudes corresponding to the
  two alternative histories of the decaying pion are:
\begin{eqnarray}
 A_i&=& <e^-p|T_R|n \nu_i> U_{ei}D( x_f-x_i,t_D-t_i, m_i ) U_{i \mu}  \nonumber \\ 
 &  & <\nu_i \mu^+|T_R|\pi^+> e^{-\frac{\Gamma_\pi}{2}(t_i-t_0)}D( 0,t_i-t_0, m_\pi )~~i=1,2 
  \end{eqnarray} 
 
 Here $U_{\alpha i}$ is the Maki-Nakagawa-Sakata (MNS)~\cite{MNS} matrix element giving the
 charged current coupling strength of the charged lepton of flavour
  $\alpha$ ($\alpha = e, \mu, \tau$) to the neutrino mass eigenstate $\nu_i$.
  The `reduced' matrix elements $<e^-p|T_R|n \nu_i>$ and $<\nu_i \mu^+|T_R|\pi^+>$
  are given by setting $U_{\alpha i} = 1$ in the charged current, so that,
  for example, $<e^-p|T|n \nu_i> = U_{e i} <e^-p|T_R|n \nu_i>$.
  Since the purely kinematical effects of non-vanishing neutrino masses are 
   expected to be negligible, then, to a very good approximation, $|\nu_1>$ and 
  $|\nu_2>$ may be replaced, in the reduced matrix elements, by the
   massless neutrino wavefunction $|\nu_0>$. The former
  are then independent of neutrino flavour. In the case of a $2 \times 2$ 
  MNS matrix for the first two generations of leptons, the unitarity of
  the matrix implies
  that it is completely defined by a single real parameter, conventionally
  chosen to be an angle, $\theta$, such that:
\begin{eqnarray}
   U_{e 1} & = &  U_{1 e}  =  U_{\mu 2} = U_{ 2 \mu} = \cos \theta \\
  U_{e 2} & = &  U_{2 e}  =  -U_{\mu 1 } =  -U_{ 1 \mu} = \sin \theta
  \end{eqnarray}
 In Eqn(2), $m_{\pi}$ and  $\Gamma_{\pi}$ are the pion pole
  mass and decay width and $D$ is the Lorentz invariant
 configuration space propagator~\cite{Feyn1,Moh1} of a neutrino or the pion. In the limit
 of large time-like separations, or of on-shell particles, appropriate
 to the experiment shown in Fig.1, 
 $D \simeq \exp[-im\Delta \tau]$ where $m$ is the pole mass of the
   particle and $\Delta \tau$ the increment of proper time corresponding to the 
  path. In the following, the additional functional dependence $\simeq (m/\Delta \tau)^\frac{3}{2}$
  of $D$ in the asymptotic region (leading to solid angle correction factors) is neglected.
  With these approximations:
 \begin{eqnarray}
 D(\Delta x,\Delta t,m) & \simeq &  \exp[-im \sqrt{(\Delta t)^2-(\Delta x)^2}] \nonumber \\
      & = &  \exp[-i m \Delta \tau] \nonumber \\  
      & \equiv & \exp[-i\Delta \phi] 
 \end{eqnarray}
 The phase increments, $\Delta \phi$, corresponding to the paths of the neutrinos and
 the pion in the amplitudes $A_i$ are:
\begin{eqnarray}
\Delta \phi_i^{\nu}& = & m_i \Delta \tau_i =\frac{m_i^2}{E_i}
 \Delta t_i = \frac{m_i^2}{P_i}L   \\
 \Delta \phi_i^{\pi}& = & m_{\pi} (t_i-t_0)
  =  m_{\pi} (t_D-t_0)-\frac{m_{\pi}L}{v_i}  \nonumber \\
 & \simeq & m_{\pi} (t_D-t_0)-m_{\pi}L\left\{1+\frac{m_i^2}{2 P_0^2}\right\}
\end{eqnarray}
  where $m_i$, $E_i$, $P_i$ and $v_i$ are the mass, energy, momentum and velocity
  of the mass eigenstate $\nu_i$ and
 \begin{equation}
 P_0 = \frac{m_{\pi}^2-m_{\mu}^2}{2 m_{\pi}}~ = ~ 29.8 \rm{MeV}
\end{equation}
  The neutrinos are assumed to follow classical rectilinear trajectories
 such that $\Delta t_i = L/ v_i = E_iL/P_i$ and the time dilatation formula
 $\Delta t =  \gamma \Delta \tau = E \Delta \tau/m$ has been used in Eqn(6).
 In Eqn(7) the neutrino velocity\footnote{Units with $\hbar = c = 1$ are used.}
 $v_i$ is expressed in terms of the neutrino mass to order $m_i^2$.
 Using Eqns(5-7), neglecting terms of O($m^4$) and higher, and replacing 
 the massive neutrino wavefunctions in the reduced matrix elements by those of massless
 neutrinos,
  the path amplitudes of Eqn(2) may be written as:
 \begin{eqnarray}
 A_i & = & <e^-p|T_R|n \nu_0> U_{e i} U_{i \mu}<\nu_0 \mu^+|T_R|\pi^+> \nonumber \\
 &   &  
 \exp[ i\phi_0-\frac{\Gamma_\pi}{2}(t_D-t_0-t_i^{fl})] 
\exp i \left[ \frac{ m_i^2}{P_0}\left(
 \frac{m_{\pi}}{2 P_0}-1\right)L \right]~~~i=1,2 
 \end{eqnarray} 
 where the neutrino times-of-flight $t_i^{fl} = t_D - t_i$ have been introduced
 and 
\begin{equation}
\phi_0 \equiv m_{\pi}(L-t_D+t_0)
\end{equation}
 Using now Eqn(1) to calculate the transition probability, and integrating over
 the detection time $t_D$~\cite{JHF}, gives, for the probability to observe the
 reaction $(\nu_1,\nu_2) n \rightarrow e^- p$ at distance $L$ from the decay point:
\begin{equation}
 P(e^-p|L) = C_N(\nu;\pi)\ \sin^2 \theta \cos^2 \theta(1-F^{\nu}(\Gamma_{\pi})\cos \phi_{12}^{\nu,\pi})
\end{equation}
where 
\begin{equation}
\phi_{12}^{\nu,\pi} = \frac{\Delta m^2}{P_0}\left(
 \frac{m_{\pi}}{2 P_0}-1\right)L = \frac{2 m_{\pi} m_{\mu}^2 \Delta m^2 L}
 {(m_{\pi}^2-m_{\mu}^2)^2} 
\end{equation}
\begin{equation}
F^{\nu}(\Gamma_{\pi})=\exp\left(-\frac{\Gamma_{\pi} m_{\pi}}{2 m_{\mu}^2}
 \phi_{12}^{\nu,\pi}\right) 
\end{equation}
 and Eqns(3) and (4) have been used. Here $C_N(\nu;\pi)$ is an $L$ independent
 normalisation factor and $\Delta m^2 \equiv m_1^2-m_2^2$. The first
 and second terms in the second member of Eqn(12) are the contributions
 of $\Delta \phi_i^{\pi}$ and  $\Delta \phi_i^{\nu}$ to the interference
 phase. The latter is a factor of two larger than in the conventional
 value~\cite{BilPont2, BiPet} $\phi_{12} = \Delta m^2 L/(2P)$.
    $\Delta \phi_i^{\pi}$ gives a numerically large
 ($m_{\pi}/2P_0=2.34$) contribution to the oscillation phase.
   
 \par The above calculation is readily repeated for the case of 
 detection of muon decay $\mu^+ \rightarrow e^+ (\nu_1,\nu_2)
 (\overline{\nu}_{1}, \overline{\nu}_{2})$
 at distance $L$ from the $\pi^+$ decay point. The phase increments
 analagous to Eqns(6) and (7) are:
\begin{eqnarray} 
\Delta \phi^{\mu}_i & = & \frac{m_{\mu}^2 L }{P_0}\left[
 1+\frac{m_i^2 E_0^{\mu}}{2 m_{\pi} P_0^2}\right]  \\
 \Delta \phi_i^{\pi(\mu)} & = & m_{\pi} (t_D-t_0)
 -\frac{m_{\pi}L}{v^{\mu}_0}\left[1+\frac{4 m_i^2 m_{\pi}^2 m_{\mu}^2}
{(m_{\pi}^2-m_{\mu}^2)^2(m_{\pi}^2+m_{\mu}^2)} \right]  
\end{eqnarray}    
where 
\[ E_0^{\mu} = \frac{m_{\pi}^2+m_{\mu}^2}{2 m_{\pi}}~~~~~~~{\rm and} 
 ~~~~~~~ 
v^{\mu}_0 = \frac{m_{\pi}^2-m_{\mu}^2}{m_{\pi}^2+m_{\mu}^2} \]
 The result found for the time-integrated decay probability is:
\begin{equation}
 P(e^+\nu \overline{\nu} |L) = C_N(\mu;\pi)(1-F^{\mu}(\Gamma_{\pi})
\sin 2 \theta \cos \phi_{12}^{\mu,\pi})
\end{equation}
where 
\begin{equation}
 \phi_{12}^{\mu,\pi} =  \frac{m_{\mu}^2 \Delta m^2}{2 P_0^3}
\left(1-\frac{ E_0^{\mu}}{m_{\pi}}\right)L = \frac{2 m_{\pi} m_{\mu}^2 \Delta m^2 L}
 {(m_{\pi}^2-m_{\mu}^2)^2} = \phi_{12}^{\nu,\pi}
\end{equation}
\begin{equation}
F^{\mu}(\Gamma_{\pi})= \exp \left(-\frac{\Gamma_{\pi}
 m_{\pi}}{(m_{\pi}^2-m_{\pi}^2)} \phi_{12}^{\mu,\pi} \right)
\end{equation}
The first
 and second terms in the second member of Eqn(17) are the contributions
 of $\Delta \phi_i^{\pi(\mu)}$ and  $\Delta \phi_i^{\mu}$ to the interference
 phase. It is interesting to note that neutrino and muon oscillation
 phases are the same for given values of $\Delta m^2$ and $L$.
 For oscillation phases $\phi_{12}^{\nu,\pi} = \phi_{12}^{\mu,\pi} =1$, the damping
  factors of the oscillation term, due to the non-vanishing pion lifetime
  take the values $F^{\nu}(\Gamma_{\pi}) = 1-1.58\times10^{-16}$ and 
$F^{\mu}(\Gamma_{\pi}) = 1-4.4\times10^{-16}$. This damping effect is thus 
  completely negligible in typical neutrino oscillation experiments with
  oscillation phases of order unity.
 
\par Formulae like Eqn(11) have been derived in a similar manner~\cite{JHF}
 for `$\overline{\nu}_{\mu} \rightarrow \overline{\nu}_e$' oscillations
 \footnote{ If the neutrinos are massive, the `lepton flavour eigenstates'
  $\overline{\nu}_e$ and $\overline{\nu}_{\mu}$ are unphysical. However
 the phrase  `$\overline{\nu}_{\mu} \rightarrow \overline{\nu}_e$ oscillations'
 is still a useful and compact way of describing experiments where
  anti-neutrinos, produced in association with a muon, give rise to a detection
  event containing an electron. The amplitudes for all physical
  processes contain, however, only the physical states $\nu_i$.}
 following muon decay at rest, 
 detected  via the process
  $(\overline{\nu}_1,\overline{\nu}_2)  p \rightarrow e^+ n$ and 
  `$\overline{\nu}_e \rightarrow \overline{\nu}_e$' oscillations following
  nuclear $\beta-$decay, detected via the same process. 
  The results found, for the time-integrated decay probabilities, are:
\begin{eqnarray}
 P(e^+n,\mu|L) & = & C_N(\overline{\nu};\mu) \sin^2 \theta \cos^2 \theta\left[1-\cos
 \frac{\Delta m^2}{P_{\overline{\nu}}}\left(
 \frac{m_{\mu}}{2 P_{\overline{\nu}}}-1\right)L\right]  \\ 
 P(e^+n,\beta|L) & = & C_N(\overline{\nu};\beta)
 \left[ \sin^4\theta + \cos^4\theta+2 \sin^2 \theta \cos^2
 \theta \cos \frac{\Delta m^2}{P_{\overline{\nu}}}\left(
 \frac{E_{\beta}}{2 P_{\overline{\nu}}}-1 \right)L \right]~ 
\end{eqnarray}
In these formulae, $ P_{\overline{\nu}}$ is the momentum of the detected
 $\overline{\nu}_1$, $\overline{\nu}_2$ , $E_{\beta}$ is the total energy release in the
 $\beta-$decay process, and damping corrections due to the finite lifetimes of
  the decaying particles are neglected.
\par Finally, in Ref.~\cite{JHF}, the cases of neutrino and muon oscillations
 following the decay in flight of ultrarelativistic pions were considered.
 The oscillation probability formulae are the same as Eqns(11) and (16)
 respectively, with the phases:
\begin{eqnarray}
\phi_{12}^{\nu,\pi}(\rm{in~flight}) & = & \frac{ m_{\mu}^2 \Delta m^2 L}
{(m_{\pi}^2-m_{\mu}^2) E_{\nu} \cos \theta_{\nu}}   \\
 \phi_{12}^{\mu,\pi}(\rm{in~flight}) & = & \frac{2 m_{\mu}^2 \Delta m^2 ( m_{\mu}^2 E_{\pi}-
m_{\pi}^2 E_{\mu})L}{(m_{\pi}^2-m_{\mu}^2)^2 E_{\mu}^2 \cos \theta_{\mu} }
\end{eqnarray}
 The neutrinos or muons are detected at distance $L$ from the decay point,
 along the direction of flight of the parent pion, with decay angles $\theta_{\nu}$
 and $\theta_{\mu}$ respectively.
 \par Because the parent pion and the daughter muon in the decay process 
 $\pi \rightarrow  \mu \nu$ are unstable particles, their physical masses
 $W_{\pi}$ and $W_{\mu}$  have distributions depending, through Breit-Wigner
 amplitudes, on their pole masses $m_{\pi}$ and $m_{\mu}$ and decay widths
 $\Gamma_{\pi}$ and $\Gamma_{\mu}$. Energy-momentum conservation in the 
  pion decay process then leads to a distribution of path amplitudes 
  with different neutrino momenta, an effect neglected in the above
  discussion. In Ref.\cite{JHF} corrections resulting from the (coherent)
 momentum smearing due to $W_{\mu}$ and (incoherent) smearing due to
 $W_{\pi}$ are calculated using a Gaussian approximation for
 the Breit-Wigner amplitudes. The resulting damping corrections
  due to $W_{\pi}$ to the
  interference terms in Eqns(11) and (16) are found to be vanishingly small.
  For $\Delta m^2 =$ (1eV)$^2$ and $L = 30$m (typical of the LNSD~\cite{LNSD}
  or KARMEN~\cite{KARMEN} experiments) the corresponding damping factors
  are found to be $1-1.3\times10^{-29}$ (for neutrino oscillations)
  and $1-5.2 \times 10^{-30}$ (for muon oscillations)~\cite{JHF}.
  The damping corrections from the variation of $W_{\mu}$ are even smaller.
  \par Corrections to Eqns(11) and (16) due to thermal motion of the decaying 
  pion and finite target and detector sizes have also been evaluated in
  Ref.\cite{JHF}. For neutrino oscillations with $\Delta m^2 =$ (1eV)$^2$
  and $L = 30$m and a room-temperature target, the damping factor of the
  interference term is found to be $1-6.7\times10^{-10}$ and the shift
  in the oscillation phase to be $1.2\times10^{-9}$rad. In summary, all
  known sources of damping of the interference terms in Eqn(10) and (15)
  are expected to be completely negligible in any forseeable neutrino
  or muon oscillation experiment.
   \par The first published calculation of the neutrino oscillation
   phase~\cite{GribPont} gave a result (only the contribution of the neutrino
   propagators was taken into account) in agreement with Eqn(6) above.
  A later calculation~\cite{FritMink} assumed instead that the phase of
  the neutrino propagator evolves as $Et$, i.e. according to the
  non-relativistic Schr\"{o}dinger equation. The assumptions were 
  also made that the different neutrino mass states are produced
   at the same time and have equal momenta. 
  The first assumption leads to the `standard' formula for the oscillation phase:
   $\phi_{12} = \Delta m^2 L/(2P)$~\cite{BilPont2, BiPet}
   which has subsequently been
   used for the analysis of all neutrino oscillation experiments.
   The Lorentz-invariant phase of Eqn(6) $\simeq m^2t/E$ evidently
   agrees with the result of Ref.\cite{FritMink} in the non-relativistic
   limit where $E \simeq m$, but such a limit is clearly inappropriate
   to describe experiments with ultra-relativistic neutrinos.
   \par  The most important difference in the treatment given in the 
   present paper to previous ones that have appeared in the literature
   is allowing the possibility for the different neutrino mass eigenstates
   to be produced at different times. Only in this way can the constraints, 
   of both space-time geometry (the detection event is at a unique space-time
    point), and exact energy-momentum conservation in the decay process~\cite{Win},
     be satisfied.
    \par The other new feature is the inclusion of the important
    contribution to the oscillation phase from the propagator of the decaying
    particle, a necessary consequence of the different production times
    of the different mass eigenstates. 
     \par In the standard treatment, the common production time of the mass 
     eigenstates follows from the assumption that a `neutrino flavour eigenstate',
  that is a coherent superposition of mass eigenstates, is produced in the decay
   of the source particle. It has been shown that the production of such states
   in pion decay is incompatible with the measured branching ratio 
  $\Gamma(\pi \rightarrow e \nu)/ \Gamma(\pi \rightarrow  \mu \nu)$ and the
  values of the  MNS matrix elements deduced from atmospheric and solar 
  neutrino oscillations~\cite{JHF1}. Actually, as first pointed out by
  Shrock~\cite{Shrock1, Shrock2}, as a consequence of the Standard 
  Model structure of the charged lepton current:
 \begin{equation}
 J_{\mu}(CC)^{lept} = \sum_{\alpha,i} \overline{\psi}_{\alpha}
  \gamma_{\mu}(1-\gamma_5)U_{\alpha i}\psi_{\nu_i}
 \end{equation}
 the different neutrino mass eigenstates are produced incoherently 
 in different physical processes. This is the basic assumption
  of the calculations presented above. Clearly, in this case, 
  the different mass eigenstates may be produced at different times 
  in the (classically) alternative histories corresponding to the
  different path amplitudes. It has also been recently demonstrated
  ~\cite{JHF2} that the factor of two difference in the contribution
 of neutrino propagation to the oscillation phase between the calculation
  presented above and the standard formula is a necessary consequence
  of the assumption of equal production times for all mass eigenstates
  and hence equal space-time (as contrasted to kinematical
 \footnote{The `kinematical' velocity, $v_{kin}$ is defined as
   $v_{kin} = p/E$.}) velocities. In the same paper ~\cite{JHF2}
   it is also shown that different kinematical hypotheses that
   lead to different, but unequal, kinematical velocities
   (equal momenta, equal energies, or exact energy-momentum
    conservation in the production process) differ in their 
   predictions for the oscillation phase only by terms of 
   O($m_i^4$), and so are all equivalent at O($m_i^2$).

\par It may be remarked that the physical interpretation of `neutrino oscillations'
  provided by the path amplitude description is different from the conventional
  one in terms of `lepton flavour eigenstates'. In the latter, the amplitudes of different
  lepton flavours in the neutrino are supposed to vary harmonically as a function of time.
  In the amplitudes for the different physical processes in the path amplitude
  treatment there is, instead, no variation of lepton flavour in the propagating
  neutrinos. 
   Only in the detection process itself, the associated MNS
   matrix elements give different coupling strengths between the 
   different neutrino mass eigenstates and the final state charged
   lepton and the interference effect occurs between the different 
   path amplitudes that is 
   described as `neutrino oscillations'. In the case of the 
   observation of the recoil muons no such `lepton flavour projection'
   occurs in the detection event,
   but exactly similar interference effects are predicted to occur.
   As previously emphasised~\cite{SWS}, the `flavour oscillations'
   of neutrinos, neutral kaons and b-mesons are just special examples
   of the universal phenomenon of quantum mechanical superposition
   that is the physical basis of Eqn(1). 
   \par The most important practical conclusion of the work presented
    here is that identical information on neutrino masses is given by
    the observation of either neutrinos or muons from pion decay at
    rest. In view of the possibility of detecting muons simply and 
    with essentially 100 $\%$ efficiency, in contrast to the tiny
    observable event rates of neutrino interactions, the recent
    indications for `$\overline{\nu}_{\mu} \rightarrow \overline{\nu}_e$' 
    oscillations with $\Delta m^2 \simeq $(1eV)$^2$ following $\mu^+$
    decay at rest~\cite{LNSD} could be easily checked by a search 
    for muon oscillations following $\pi^+$ decay at rest. For  
    $\Delta m^2 \simeq $(1eV)$^2$, the first absolute maximum of the
    interference term in Eqn(16) occurs at $L \simeq 8$m. Note that
    Eqn(16) is valid for any muon detection process that does not 
    distinguish between events where $\nu_1$ or $\nu_2$ are produced.  
 
\pagebreak

\newpage
\vspace*{4cm}
\begin{figure}[htbp]
\begin{center}
\hspace*{-0.5cm}\mbox{
\epsfysize15.0cm\epsffile{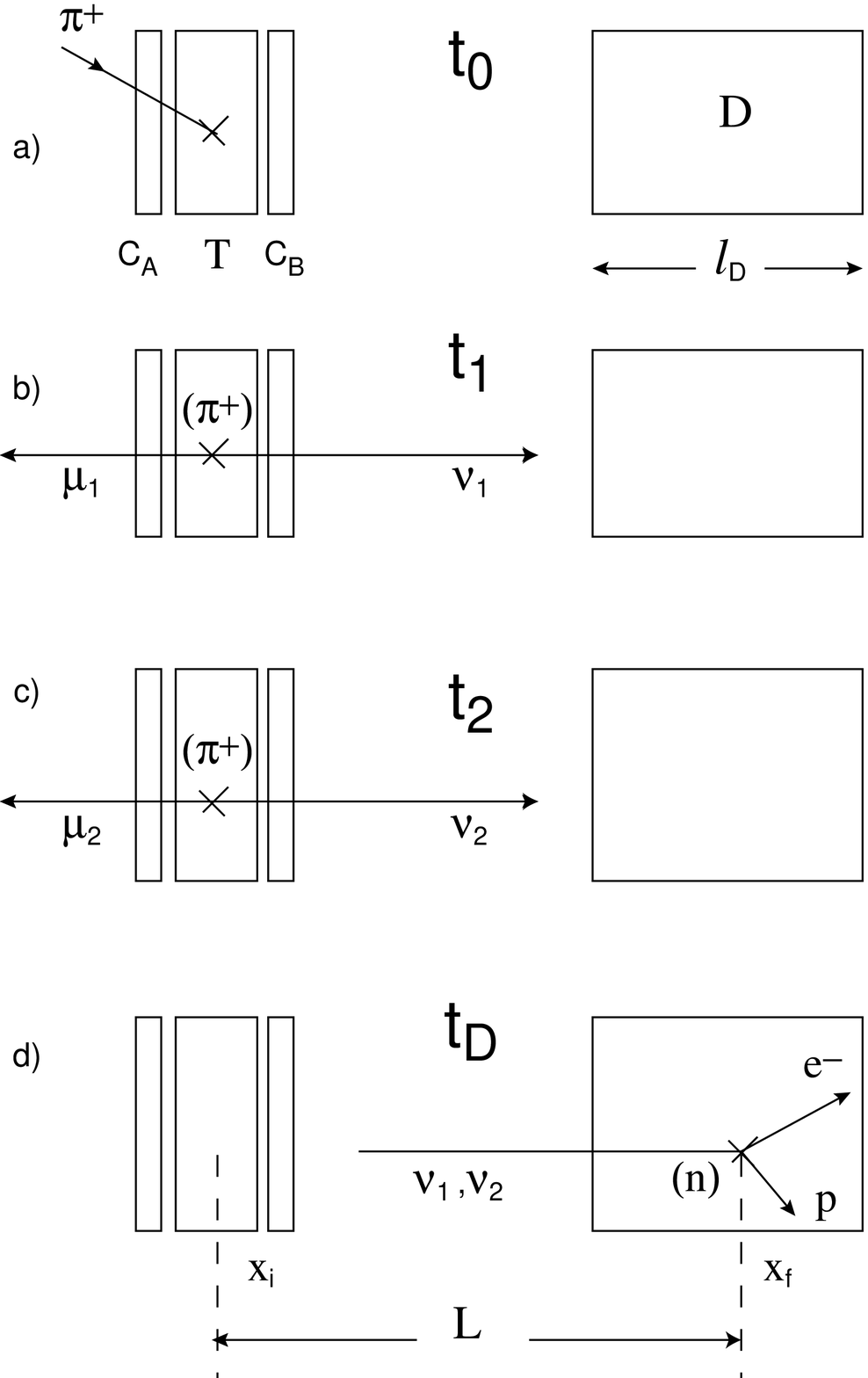}}
\caption{The space-time description of an experiment in which neutrinos
 produced in the processes $\pi^+ \rightarrow \mu^+ (\nu_1, \nu_2)$ are detected at
  distance, $L$, via the processes $(\nu_1, \nu_2) n \rightarrow e^- p$ (see text).}
\label{fig-fig1}
\end{center}
 \end{figure}
\end{document}